\documentclass[12pt,preprint]{aastex}
\usepackage[latin9]{inputenc}
\usepackage{esint}

\makeatletter

\newcommand{\simlt}{\lower.5ex\hbox{$\; \buildrel < \over \sim \;$}}
\newcommand{\simgt}{\lower.5ex\hbox{$\; \buildrel > \over \sim \;$}}

\makeatother

\begin{document}

\title{Sub-photospheric, radiation mediated shocks in GRBs: Multiple shock
emission and the Band spectrum}

\author{Shai Keren$^{1}$ and Amir Levinson$^{1}$}
\altaffiltext{1}{School of Physics \& Astronomy, Tel Aviv University, Tel Aviv 69978,
Israel; Levinson@wise.tau.ac.il}

\begin{abstract}
We compute the time integrated, thermal emission produced by a series of radiation mediated shocks that 
emerge from the photosphere of a GRB outflow.   We show that for a sufficiently broad distribution of 
shock strengths, the overall shape of the time integrated spectral energy
distribution below the peak is a power law, $\nu E_\nu\propto \nu^\alpha$, with a 
slope $1<\alpha<2$.   A substructure in the SED can also be produced 
in this model for certain choices of the shock train distribution.   In particular, 
we demonstrate that our model can reproduce the double-peak SED observed in some bursts, 
in events whereby a strong shock is followed by a sequence sufficiently weaker ones. 
\end{abstract}

\section{Introduction}

The origin of the prompt emission observed in gamma ray bursts (GRBs) is yet an open issue.  The emitted spectrum, although exhibits notable 
variations from source to source, can be generally described by a Band function (Band et al. 1993), with some exceptions, e.g., GRB 090902B. A common interpretation of the characteristic Band spectrum observed in many GRBs is synchrotron emission of nonthermal electrons 
accelerated at collisionless shocks.  However, the synchrotron model has difficulties accounting for some common properties exhibited by the GRB population, specifically, the clustering of peak energies around 1 MeV, the hardness of the spectrum below the peak, and the high efficiencies inferred from the observations  (e.g., Crider et al. 1997; Preece et al. 1998; Eichler \& Levinson 2000; Ghirlanda et al. 2003; Beloborodov 2013). 

The difficulties with the synchrotron model (and more generally, optically thin emission models), and the recent detections of some GRBs 
with a prominent thermal component (e.g., GRB 090902B), or multiple peaks (e.g., GRB 110721A, GRB 120323A), have motivated 
reconsideration of photospheric emission (Eichler \& Levinson 2000; Peer, et al. 2006, Ryde \& Peer 2009; Peer, et al. 2012; Giannios 2012; Beloborodov 2013).   On theoretical 
grounds, it has been argued (Levinson 2012) that a significant dissipation of the bulk energy of a GRB outflow just below the photosphere, either by internal shocks (Eichler 1994; Bromberg et al. 2011) or collimation 
shocks (Lazzati et al. 2009; Morsony et al. 2010),  is a natural expectation in case of a hydrodynamic flow.   We emphasize that this mechanism may also be
relevant for magnetically extracted MHD flows if conversion of magnetic energy to kinetic energy occurs below the 
photosphere, as proposed recently in  Levinson \& Begelman (2013, but c.f., Bromberg et al. 2014).   In general, the structure and emission of sub-photospheric shocks 
are vastly different than those of collisionless shocks that can only form above the photosphere, where the Thomson optical depth is sufficiently 
small (Levinson and Bromberg 2008; Katz et al. 2010, Budnik et al. 2010, Bromberg et al. 2011, Levinson 2012).    Shocks that form below the photosphere are mediated by Compton scattering of  radiation advected into the shock by the upstream fluid.  The characteristic length scale of
the shock transition layer is a few Thomson lengths,  vastly larger than any kinetic scale involved, hence these shocks cannot accelerate particles to nonthermal energies.   The time integrated spectrum emitted by a shock that emerges from the photosphere  consists of two main components; a
quasi thermal component contributed by the hot gas downstream of the shock, and a hard component which is produced inside the shock transition layer.  The hard component extends up to an energy of about  $\gamma_um_ec^2$, as measured  in the shock frame, where $\gamma_u$ is the Lorentz factor of the upstream flow with respect to the shock frame (some examples are exhibited in Budnik et al. 2010, and   Bromberg et al. 2011).

In  a previous paper (Levinson 2012) we analyzed the properties of a sub-photospheric shock that forms in a GRB jet, and computed the time integrated spectrum emitted by a single shock event.    Only the contribution from the downstream region, which dictating the shape of the spectrum below the peak, was accounted for, since the computation of the hard component produced inside the shock transition layer requires sophisticated 
Monte-Carlo simulations, that are still in progress.    It has been shown that the spectrum emitted by a single shock exhibits a prominent thermal peak, with the location of the peak depending on the velocity profile of the shock and the specific entropy of the upstream flow.   
In this paper we generalize the analysis to multiple shock emission.   We show that the time integrated emission from several shocks can mimic a Band spectrum, and can also account for multiple peaks in certain cases, like those seen in  GRB 110721A and  GRB 120323A.

\section{Structure of radiation mediated shocks}
The general structure of a radiation mediated shock (RMS) consists of a shock transition layer inside which the upstream flow is decelerated via the 
Compton drag exerted on it by counter-streaming photons, an immediate post shock region where the decelerating
flow reaches its terminal velocity and the radiation field becomes isotropic (in the frame of the shocked fluid), and a thermalization layer
in which the photon density gradually increases, and the temperature decreases, owing to free-free and double Compton 
emissions, until a full thermodynamic equilibrium is established.     Under the conditions anticipated in many GRBs, photon advection by the upstream
flow dominates over photon production inside and just behind the shock transition layer (see Bromberg et al. 2011, and Levinson 2012 for 
more details).   Then, the temperature $T_s$ and photon density $n_{rs}$ in the 
immediate post shock region can be readily computed using the shock jump conditions.   The temperature profile in the thermalization 
layer is computed by employing a transfer equation that incorporates photon production processes (and adiabatic cooling when important),
whereby $T_s$ is used as a boundary condition (Levinson 2012). 

\subsection{Jump conditions}

For a planar shock propagating in a homogeneous medium consisting
of protons, electrons and seed radiation, the fluid parameters in the immediate downstream
are  obtained from the jump conditions 
\begin{eqnarray}
n_{bu}u_{u}=n_{bs}u_{s}, \label{eq:bar-density}\\
n_{ru}/n_{bu}=n_{rs}/n_{bs}, \label{eq:phot-density}\\
w_{u}u_{u}^{2}+p_{u}=w_{s}u_{s}^{2}+p_{s}, \label{eq:jump2}\\
w_{u}\gamma_{u}u_{u}=w_{s}\gamma_{s}u_{s}, \label{eq:jump3}
\end{eqnarray}
where subscripts $u$ and $s$ refer to the upstream and immediate downstream
values of the fluid parameters, respectively,  $n_{b}$ denotes the
baryon density, $n_{r}$ the photon density, $w$ the specific enthalpy, $p$ pressure, $u$ the fluid
4-velocity with respect to the shock frame, and $\gamma=\sqrt{1+u^2}$  the corresponding
Lorenz factor.  Equation (\ref{eq:phot-density}) ignores photon production inside the shock.
As explained above, this is a good approximation for sub-photospheric shocks that form in GRB outflows. 
For the situations considered in this paper, the pressure,
when important, is always dominated by radiation, viz., $p=p_r$, where $p_r$ denotes the radiation pressure. Thus, the specific
enthalpy, both upstream and downstream of the shock transition layer,
can be approximated as $w=n_{b}m_{p}c^{2}+4p_{r}$.

The average photon energy in the immediate post shock region is  $\langle h\nu\rangle_s=e_{rs}/n_{rs}$.
If the radiation in the immediate downstream is fully Comptonized, then a local Wien spectrum is established, with a temperature of
$kT_s=\langle h\nu\rangle_s/3=p_{rs}/n_{rs}$.   The average number of scatterings of a photon downstream is 
$N_{sc}={\rm min}\{\tau^2,\tau/\beta_s\}$.   The Compton parameter is given by $y_c=(4kT_s/m_ec^2)N_{sc}$.  A photon
of initial energy $h\nu_0$ will join the peak provided $y_c>\ln (3kT_s/h\nu_0)$.  With $\beta_s=1/3$ we then estimate that 
a Wien spectrum will be established downstream at a distance of $\tau\ge(m_ec^2/12kT_s)\ln(3kT_s/h\nu_0)= $ a few, from the shock front.
 In terms of  the dimensionless quantities $\tilde{n}= n_{ru}/n_{bu}$ 
and $\pi_{s}=p_{s}/(n_{bu}m_pc^{2})$, the Compton temperature downstream can be expressed as
\begin{equation}
kT_{s}=\tilde{n}^{-1}m_{p}c^{2}\pi_{s}(u_{s}/u_{u})\label{eq:kTs}
\end{equation}
where Equations (\ref{eq:bar-density}) and (\ref{eq:phot-density}) have bee employed.

Equations (\ref{eq:bar-density})-(\ref{eq:jump3}) can be solved analytically in the ultra-relativistic and non-relativistic limits. 
For the cases considered here the ratio of radiation pressure and rest mass energy density upstream
of the shock, $\tilde{p}=p_{ru}/(n_{bu}m_{p}c^{2})$, is typically small.  In the ultra-relativistic limit we can therefore 
set  $p_{u}=0$ in Equations (\ref{eq:jump2}) and (\ref{eq:jump3}). The jump  conditions then yield 
\begin{eqnarray}
u_{s}=1/\sqrt{8}, \label{eq:beta_s-UR}\\
e_{rs}=3p_{rs}=2n_{bu}u_{u}^{2}m_{p}c^{2}. \label{eq:e_rs_UR}
\end{eqnarray} 
From  Equation (\ref{eq:e_rs_UR}) we obtain $\pi_s=2u_u^2/3$.  Substituting 
the latter results into Equation (\ref{eq:kTs}), and denoting $\tilde{n}=10^4\tilde{n}_4$, yields 
\begin{equation}
kT_{s}\simeq23\tilde{n}_{4}^{-1}\gamma_{u}\beta_{u}\, {\rm keV}.\label{eq:kTs RRMS}
\end{equation}

In the nonrelativistic limit, the radiation pressure upstream cannot be ignored.  Solving the jump 
conditions in terms of $\tilde{p}=p_{ru}/(n_{bu}m_pc^2)$, we obtain
\begin{eqnarray}
\beta_{s}=\frac{\beta_{u}}{7}(1+8\tilde{p}/\beta_{u}^{2}),\\
e_{rs}=3p_{rs}=\frac{18}{7}n_{bu}\beta_{u}^{2}m_{p}c^{2}(1-\tilde{p}/6\beta_{u}^{2}), \label{eq:e_rs_NR}
\end{eqnarray}
where $\beta=u/\gamma\simeq u$ is the 3-velocity.   Substituting the latter relations into Equation (\ref{eq:kTs}) we have,
\begin{equation}
kT_{s}\simeq10\tilde{n}_{4}^{-1}\beta_{u}^{2}(1+8\tilde{p}\beta_{u}^{-2})\, {\rm keV}. \label{eq:kTs NRRMS}
\end{equation}

\subsection{Upstream conditions in  GRB outflows}
As shown above, the temperature $T_s$ just downstream of the shock depends on the photon-to-baryon ratio in the unshocked 
flow,  $\tilde{n}=n_{ru}/n_{bu}$.    This ratio depends, in turn, on the injection and dissipation process near the central 
engine.   A simple way to estimate it is to assume that the flow becomes adiabatic above some radius $R_0=10^7R_7$ cm, at which 
its Lorentz factor is $\Gamma_0\simgt1$.   
For simplicity, we suppose that the fireball is sufficiently opaque, such that its photosphere is located in the coasting 
region, where the bulk Lorentz factor is given by $\Gamma\simeq\eta\equiv L_{iso}/(\dot{M}_{iso}c^2)$ in terms of the 
isotropic equivalent luminosity, $L_{iso}=10^{52}L_{52}$ erg s$^{-1}$, and isotropic mass loss rate $\dot{M}_{iso}$.  It 
can then be shown that $\tilde{n}=10^{2.5}\Gamma L_{52}^{-1/4}R_7^{1/2}\Gamma_0^{-1/2}$ (Levinson 2012). 
From Equation (\ref{eq:kTs}) we then obtain
\begin{equation}
kT_{s,ob}=\Gamma kT_{s}\simeq 3L_{52}^{1/4}R_{7}^{-1/2}\Gamma_{0}^{1/2}\pi_{s}(u_{s}/u_{u})\quad {\rm MeV}.\label{eq:kTs_full}
\end{equation}
Note that observed temperature, $T_{s,ob}$, is independent of the bulk Lorentz factor $\Gamma$. 
In the ultrarelativistic limit $\pi_s(u_s/u_u)=u_u/(3\sqrt{2})$, and the letter equation yields $kT_{s,ob}\simeq700 u_u$ keV for our canonical 
choice of parameters, $L_{52}^{1/4}R_{7}^{-1/2}\Gamma_{0}^{1/2}=1$.  For intermediate values of $u_u$ 
Equations (\ref{eq:bar-density})-(\ref{eq:jump3}) must be solved numerically to obtain the downstream values $\pi_{s}$ and $u_s$.
Figure 1 presents numerical solutions for the downstream temperature
in the immediate post-shock region for a range of upstream velocities
$u_{u}$, in the regime where $\tilde{p}\ll1$. 

We note that direct emission of the radiation advected from the central engine by the unshocked flow will also contribute to the observed signal. However,
it is expected to be negligible in cases where the photosphere is located well above the coasting radius, as the radiative efficiency is
very small (see, e.g., Eq. (22) in Levinson 2012).  Thus, in general, additional dissipation is needed just below the photosphere in order to 
convert a significant fraction of the bulk energy into gamma rays.  Most recent models (e.g., Peer et al. 2006; Giannios 2012, Beloborodov 2013; Vurm et al. 2013) invoke some unspecified dissipation mechanism.     Here we propose that sub-photospheric dissipation is accomplished via RMS.

\begin{figure}[ht]
\centering
\includegraphics[width=12cm]{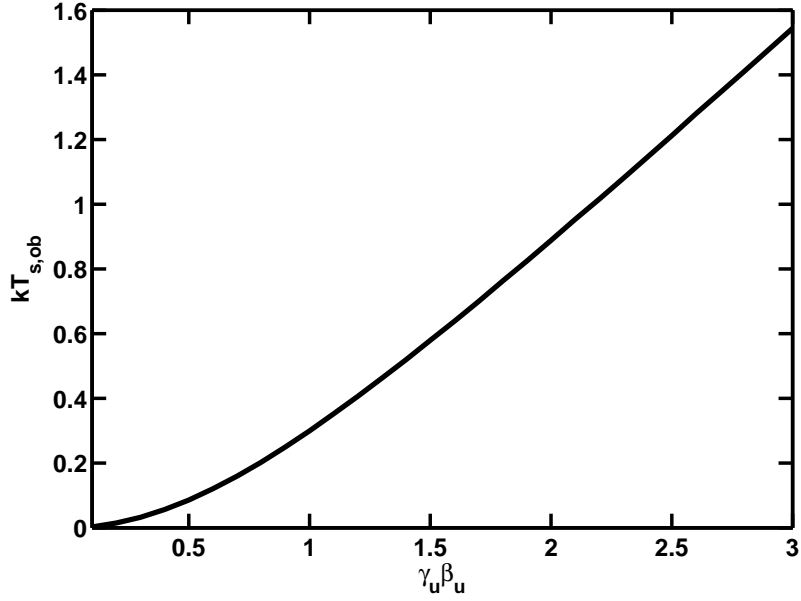}
\caption{\label{fig:f1}Downstream temperature profile as a function of the upstream four-velocity. }
 \end{figure}

\subsection{Thermalization and cooling behind the shock}

In the thermalization layer, photons are being produced by free-free and double Compton emissions, with specific rates $\dot{n}_{ff}$
and $\dot{n}_{DC}$, given explicitly in Levinson (2012).   The change in the photon density $n_{r}$
is given to a good approximation by
\begin{equation}
\partial_{\mu}(n_{r}u_s^{\mu})=\dot{n}_{ff}+\dot{n}_{DC},\label{eq:photon_generation}
\end{equation}
where $u_s^{\mu}$ is the four-velocity of the shocked gas.  The latter equation assumes that photon transport is dominated by convection (rather than diffusion), as anticipated in the downstream of a relativistic shock.

In case of GRBs the shock propagates in a relativistically expanding flow.    For simplicity, we consider a
stationary, conical unshocked flow with a proper density $n_{bu}(r)\propto r^{-2}$ and a constant bulk Lorentz factor $u^0=\Gamma$.  
The optical depth of the unshocked flow is given by $\tau(r)=\int_{r}^{\infty}\sigma_Tn_{bu}(r)dr/\Gamma$, whereby $\tau(r_{ph})=1$ at
the photospheric radius $r_{ph}$.    Suppose now that 
a shock has formed at some radius $r_0<r_{ph}$, with a corresponding optical depth $\tau_0\equiv \tau(r_0)>1$.  As the shock propagates outwards, it accumulates hot plasma in the shocked layer behind it.   The temperature of the shocked layer declines during the course of its 
evolution, owing to adiabatic cooling and photon production.    Consider a fluid shell that crossed the shock at some radius $r_s$, where $r_0<r_s<r_{ph}$.   Its pressure $p_s(r_s)$, photon density $n_{rs}(r_s)$, and temperature $T_s(r_s)$ can be computed using the local jump conditions at $r=r_s$, where $n_{bu}=n_{bu}(r_s)$, $n_{ru}=n_{ru}(r_s)$, etc.    Now, the change in the photon density $n_r(r)$ of that shell 
during its propagation from $r_s$ to $r>r_s$ is given by
\begin{equation}
\frac{c}{r^2}\frac{d}{dr}(r^2 n_{r}\Gamma)=\dot{n}_{ff}+\dot{n}_{DC}.\label{eq:photon_generation_2}
\end{equation}
where Equation (\ref{eq:photon_generation}) has been employed.   Behind the shock the flow is adiabatic, hence the equation of state
 $e_r\propto n_{b}^{4/3}$ applies.   We can then write  $e_{r}(r)=e_{rs}(r_{s}/r)^{8/3}$ with $e_{rs}(r_s)=3\pi_s(r_s)m_pc^2n_{bu}(r_s)$.    On scales smaller
than the thermalization length the trapped radiation has a Wien spectrum,
with $n_{r}=e_{r}/3kT$.   Substituting into Equation (\ref{eq:photon_generation_2}) and transforming from $r$ to the variable $\tau(r)$, one arrives at (see  Levinson 2012 for details),
\begin{equation}
\frac{d\tilde{T}}{d\tau}=\frac{2\tilde{T}}{3\tau}+A\tilde{T}^{3/2}[(\tau_{s}/\tau)^{2/3}+\kappa\tilde{T}^{3/2}],\label{eq:T_Profile}
\end{equation}
where $\tilde{T}(\tau)=T(\tau)/T_{s}(\tau_s)$, and the coefficients $A$ and $\kappa$ are given explicitly in Levinson (2012)
in the ultra-relativistic and non-relativistic limits.   Equation  (\ref{eq:T_Profile}) governs the evolution 
of the temperature of a shocked fluid shell as it propagates from its formation radius $r_s$ ($\tau_s>1$) to 
$r>r_s$ ($\tau<\tau_s$).  The first term on the right hand side accounts for
adiabatic cooling, and the second therm for photon production via free-free and double Compton emissions. 

Let $T_{ph}(\tau_s)$ denotes the temperature of a shell that crossed the shock at some optical depth $\tau_s>1$ as it reaches the
photosphere.  Then,  $T_{ph}(\tau_{s})=T_{s}(\tau_s)\tilde{T}(\tau=1)$.  For  a given $\tau_{s}$, 
$T_{s}(\tau_s)$ is computed by employing Equation (\ref{eq:kTs_full}), upon numerically solving the jump 
conditions (\ref{eq:bar-density})-(\ref{eq:jump3}) to obtain $u_s$ and $\pi_s$.  The photospheric value 
$\tilde{T}(\tau=1)$ is computed by integrating Equation (\ref{eq:T_Profile})
from $\tau=\tau_{s}$ to $\tau=1$, subject to the boundary condition $\tilde{T}(\tau=\tau_{s})=1$, whereby $T_{ph}(\tau_{s})$ is obtained. 
The photospheric temperature $T_{ph}(\tau_{s})$ is the key quantity that determines the shape of the emitted spectrum.

\section{Shock breakout and emission}
In this section we compute the time integrated spectrum emitted from the photosphere of a GRB outflow by a shock train produced via
overtaking collisions of several shells.  We shall consider cases whereby the shocks are well separated, in the sense that the upstream conditions 
of each shock are determined by the unshocked bulk flow\footnote{A more complicated situation can be envisaged, wherein a second shock forms in the downstream region of a leading shock.}. This simplified situation is sufficient, in our opinion, to elucidate the main features of the emitted spectrum.  Since the multiple shock emission is essentially a superposition of single shock spectra, we shall begin by giving a brief account of the single shock model developed in Levinson (2012). 

\subsection{\label{sec:sing-shock}Single shock emission}

The integrated spectrum emitted following shock breakout has two main components: a hard (non-thermal) component which is produced inside the shock transition layer, and extends from the thermal peak up to $\Gamma\gamma_{u}m_{e}c^{2}$ in the observer frame, and a thermal component
which is emitted, subsequently, by the hot gas downstream of the shock.     Here, we compute only the thermal component.   The calculations of the nonthermal spectrum are far more involved, and are still underway.
 
Once the shock breaks out of the photosphere, an
observer will start receiving radiation from the shocked shells. Each
shell reaches the photosphere with a temperature given by $T_{ph}(\tau_s)=T_s(\tau_s)\tilde{T}(\tau=1)$, where $\tilde{T}(\tau=1)$ is a solution
of Equation (\ref{eq:T_Profile}), as explained above.   The temperature $T_s(\tau_s)$ depends on the local shock velocity, $u_{sh}(\tau_s)$ (see Equation (\ref{eq:kTs}) with $u_u=u_{sh}$).  Thus, a model for the shock dynamics is needed.   Since the details of the shock mechanism are uncertain, we shall invoke a prescribed velocity profile, $u_{sh}(\tau)$. 

\begin{figure}[ht]
\centering
\includegraphics[width=12cm]{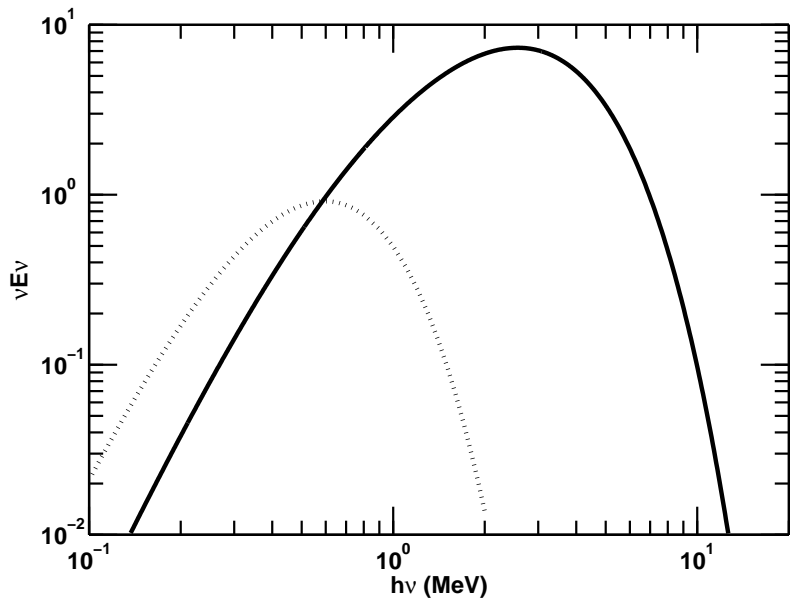}
\caption{\label{fig:f2}  Time integrated spectral energy distribution (TSED) emitted by a single shock moving at a constant
velocity, $u_{sh}=2$ (solid line), and by a decelerating shock having a velocity profile 
$u_{sh}(\tau)=2(\tau/\tau_0)^{1/2}$, with $\tau_0=10$ (dotted line).}
 \end{figure}

If the Compton parameter is sufficiently large, then the radiation trapped in the shell has a Wien spectrum.  For those shells that form 
not deep enough below the photosphere the location of the peak should be unaltered, but the spectrum below the peak is expected to be somewhat softer.    The results presented in the next section depend weakly on these details (but see further discussion there), and so we will adopt a Wien spectrum for all shells.  At the photosphere,
\begin{equation}
I_{\nu}(t=x_{s}/c)=\frac{e_{r}(r_{ph})c}{24\pi}\left(\frac{h}{kT_{ph}(\tau_{s})}\right)^{4}\nu^{3}e^{-h\nu/kT_{ph}(\tau_{s})}
\end{equation}
 where $e_{r}=3p_{d}(r_{s}/r)^{8/3}$, and $t$ is the time at which the
shell has reached the photosphere. The time integrated spectral energy
distribution (TSED) is given by
\begin{equation}
\nu E_{\nu}=\int 4\pi r_{ph}^{2}\nu I_{\nu}(t,r_{ph})dt
\propto\int_{1}^{\tau_{0}}\left(\frac{h\nu}{kT_{ph}(\tau_{s})}\right)^4\tau_{s}^{-8/3}e^{-h\nu/kT_{ph}(\tau_{s})}d\tau_{s},\label{eq:SED-single}
\end{equation}
where $\tau_{0}$ corresponds to the radius $r_0$ at which the shock was
initially formed. The integrated emission exhibits a roughly thermal
spectrum (Fig. \ref{fig:f2}), with a peak energy $h\nu_{peak}\simeq3kT_{s,ob}$, where
$T_{s,ob}$ is given by Equation (\ref{eq:kTs_full}). The portion
of the spectrum below the peak is much harder than that of a typical
Band spectrum. This result reflects a generic shortage in production
of photons by sub-photospheric shocks. A detailed discussion about
the spectrum emitted from a single shock, including relevant figures,
can be found in (Levinson 2012).

\subsection{Multiple shock emission}

We consider the time integrated emission from multiple shocks with a specified distribution of properties (e.g., formation radii, velocity profiles, etc.).
Given the assumption that the shocks are well separated, the total emission from the shock train is simply the sum 
of single-shock spectra taken over the entire multiple shock ensemble. 
For each shock we specify its formation depth $\tau^k_0$ and velocity profile $u^k_{sh}(\tau)$, where $k=0,1,..N_{sh}-1$,
and compute the TSED using Equation (\ref{eq:SED-single}) with 
absolute normalization.  We then sum up all contributions to obtain the overall TSED: $\nu E_{\nu}=\Sigma_{k=0}^{N_{sh}-1}\nu E^k_{\nu}$.

\begin{figure}[ht]
\centering
\includegraphics[width=16cm]{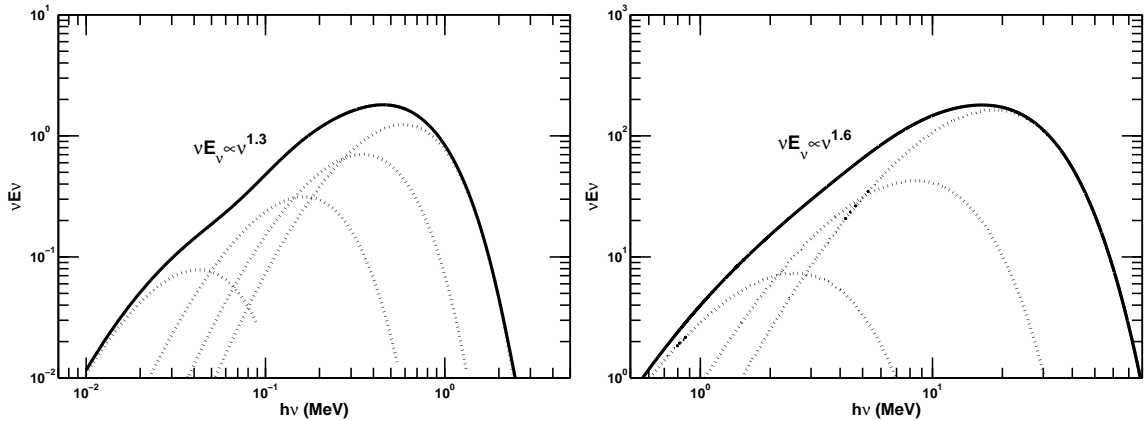}
\caption{\label{fig:f3} Left panel: the TSED emitted by a train of non-relativistic RMS with evenly distributed 4-velocities, $u^k_{sh}=0.2,0.4,0.6, 0.8$ ($k=0,1,2,3$), for our canonical model ($R_7=L_{52}=\Gamma_0=1$).  The contribution of each shock in the ensemble is delineated by the dotted curves. 
Right panel: same, but for a train of relativistic RMS with
4-velocities $u^k_{sh}=2,5,10$ ($k=0, 1, 2$).}
 \end{figure}

Some basic properties of the TSED can be readily understood using the following heuristic argument: consider an ensemble of 
non-relativistic shocks with a uniform velocity distribution.    From Equation (\ref{eq:e_rs_NR}) it is seen that
the energy dissipated behind a single shock moving at a constant velocity $u_{sh}\simeq \beta_{u}$ scales roughly as
$\beta_{u}^{2}$.   From  Equation (\ref{eq:kTs NRRMS}) it is seen that the peak energy also scales like $\beta_{u}^{2}$.
Consequently, for a uniform distribution of shock velocities we anticipate $\nu E_{\nu}\propto \nu$.  Likewise,
in the ultra-relativistic regime the dissipated energy scales like
$u_{sh}^{2}$ (Equation (\ref{eq:e_rs_UR})), and the peak energy
scales like $u_{sh}$ (Equation (\ref{eq:kTs RRMS})), hence
we expect $\nu E_{\nu}\propto \nu^{2}$. 
These trends are clearly seen in figure \ref{fig:f3}, where the TSED emitted from a train of
mildly relativistic (left panel) and ultra-relativistic (right panel) shocks is exhibited. 

\begin{figure}[ht]
\centering
\includegraphics[width=12cm]{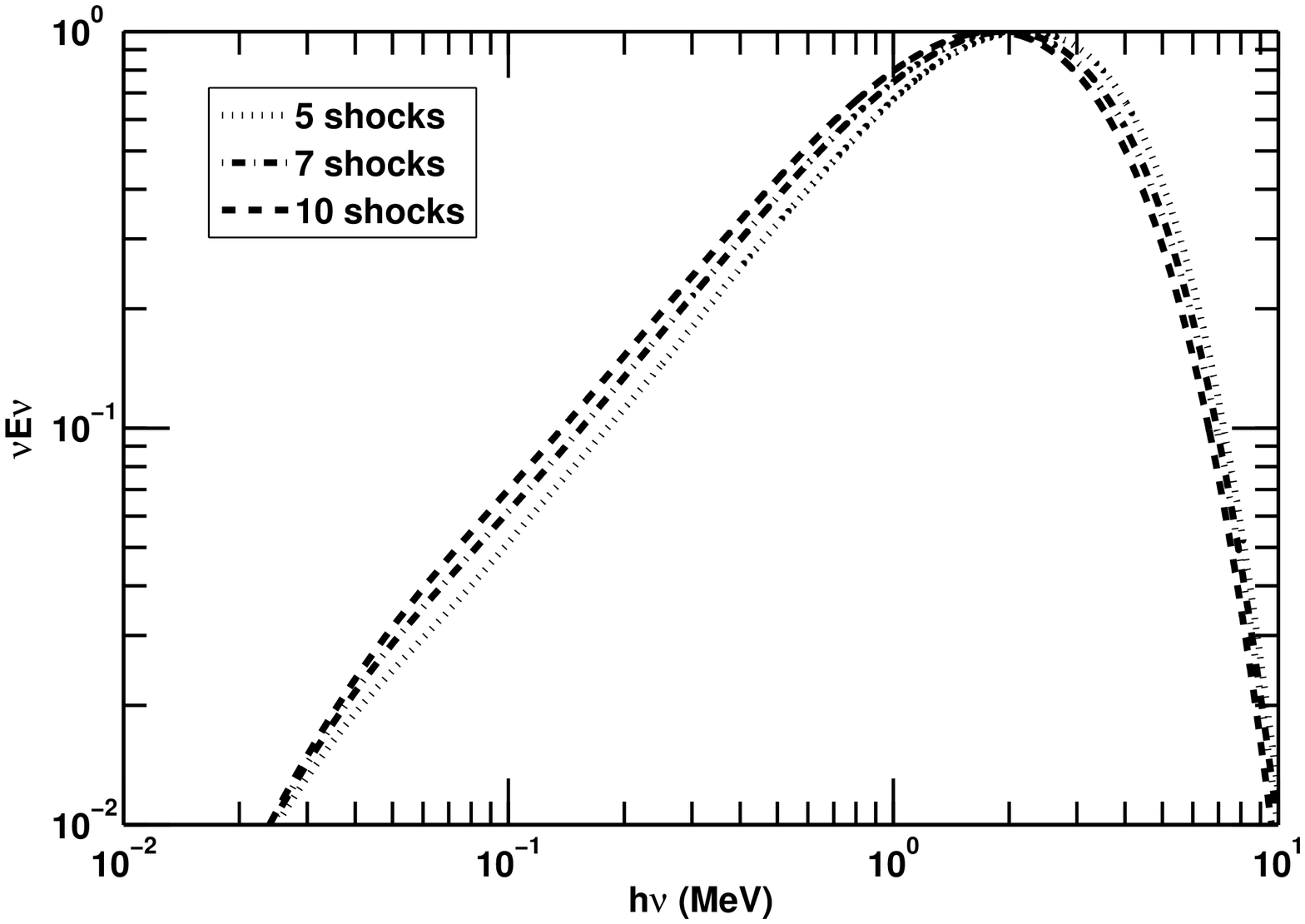}
\caption{\label{fig:f4}A comparison of the TSEDs produced by shock trains with different number $N_{sh}$ of shocks, as indicated.
The shock 4-velocities, in each case, are constant (uniform dissipation) and 
lie the range $0.2\le u_{sh}^k \le2$, with equal spacing on a logarithmic scale.  The formation depth is the same for all shocks, $\tau^k_0=10$.}
\end{figure}

\begin{figure}[ht]
\centering
\includegraphics[width=12cm]{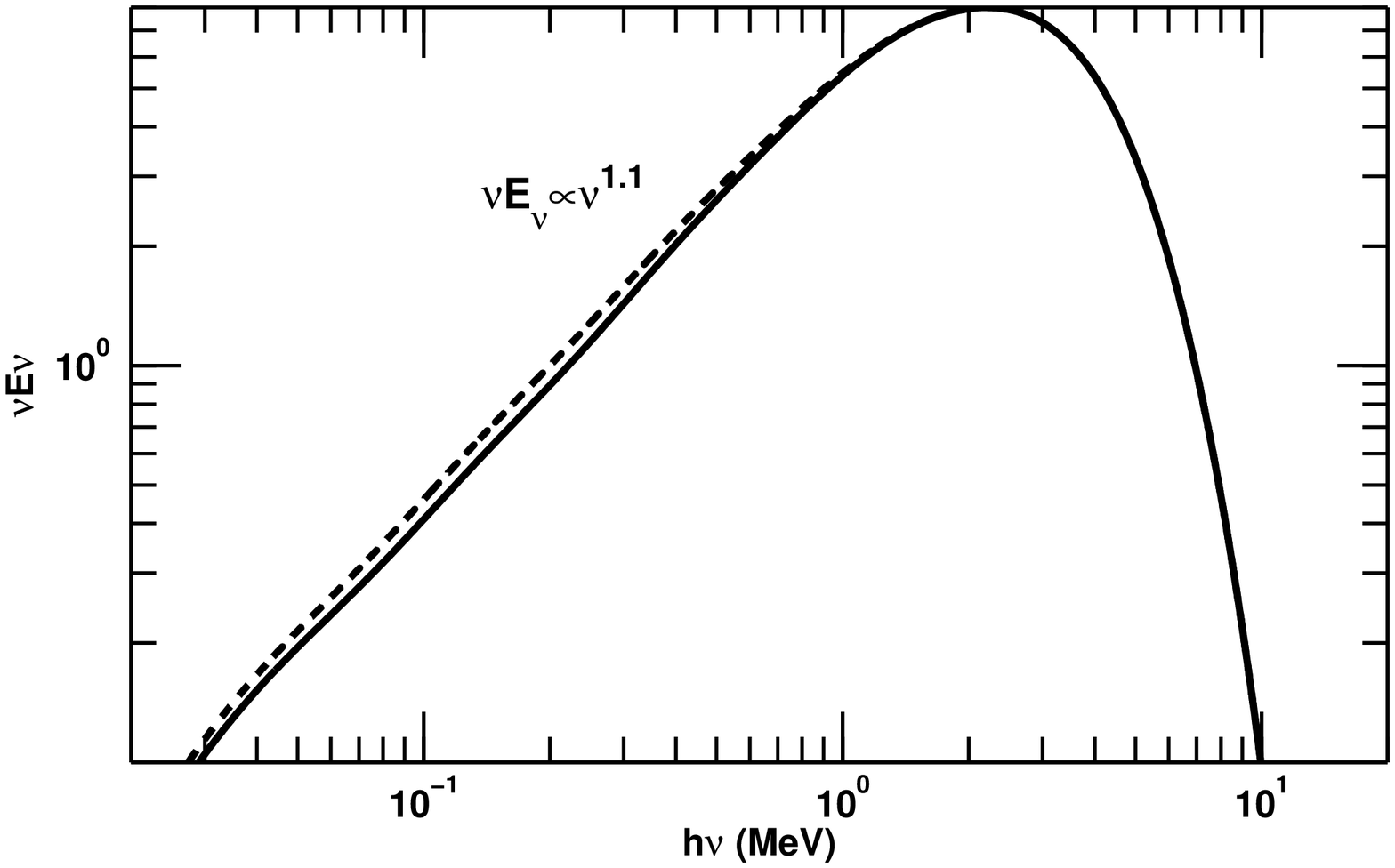}
\caption{\label{fig:f5} TSEDs produced by 5 shocks, each moving at a constant velocity, with $0.2\le u_{sh}^k \le2$ equally spaced on a logarithmic scale. The solid line corresponds to an ensemble of shocks that formed at the same depth, $\tau_0^k=10$ for $k=0,1,2,3$ (same as in Fig 4). The dashed line corresponds to a case where each shock was formed at a different depth, with $\tau^k_0$ in the range $2-100$.}
\end{figure}

Figure \ref{fig:f4} depicts the dependence of the TSEDs on the number of shocks contributing to the total emission.  
In this particular example we assume a uniform dissipation along each shock's trajectory (Levinson 2012). To be concrete, 
all shocks propagate at a constant velocity, although with different values, as indicated in the figure. 
As seen, the overall shape of the TSED below the peak is a rough power law, weakly dependent on the number of shocks, as well 
as on the distribution of shock formation radii, as illustrated in figure \ref{fig:f5}.  The 
reason is that, in case of a uniform dissipation the dominant contribution to the time-integrated emission from each 
shock comes from regions just beneath the photosphere (see LE12 for details). This may no longer be true in case of 
a more complex shock dynamics.  
Some wiggly structure may appear when the number of shocks contributing to the emission is small (3 or 4), as discussed below.

\begin{figure}[ht]
\centering
\includegraphics[width=12cm]{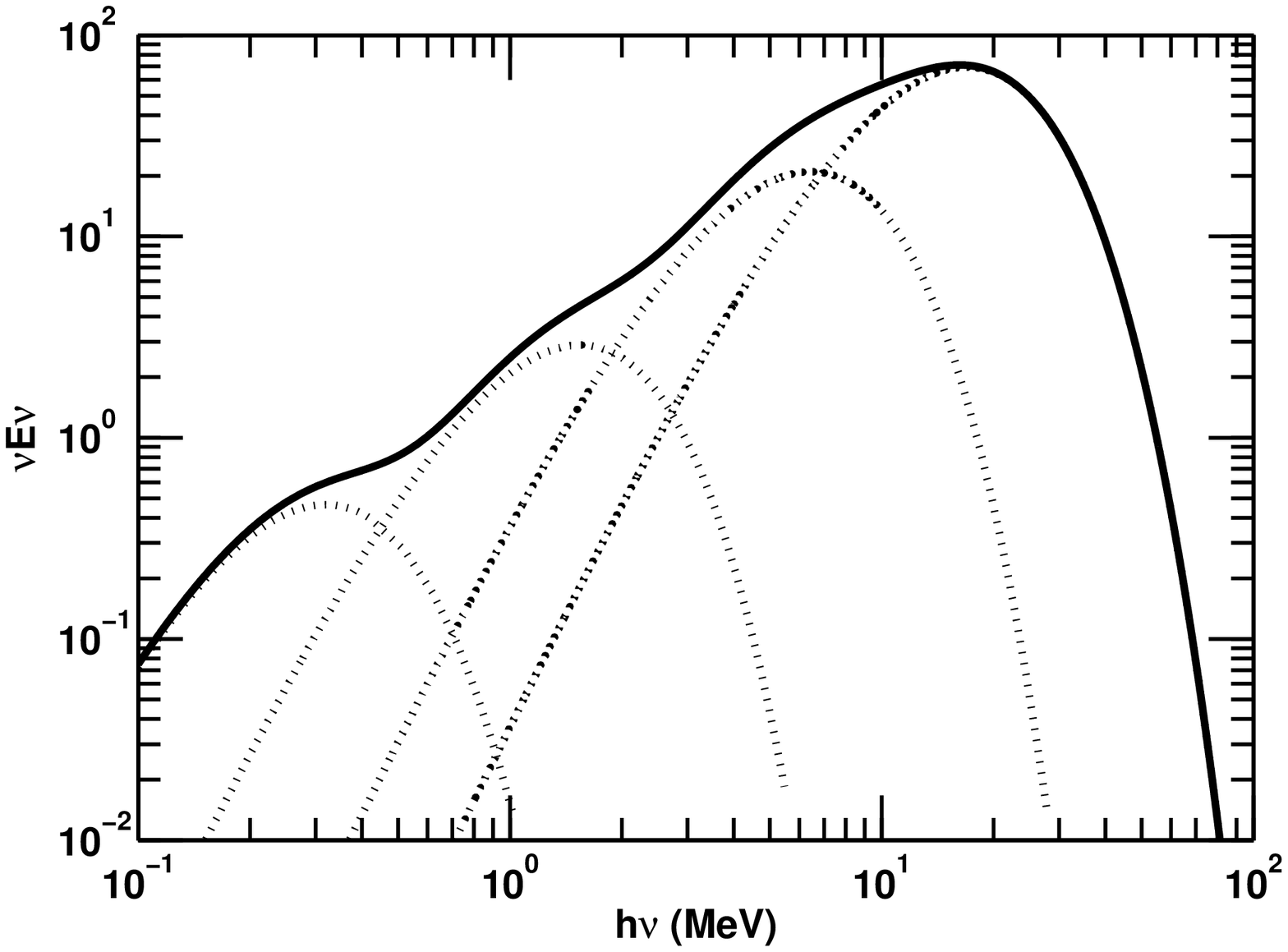}
\caption{\label{fig:f6}TSED produced by a train of $N_{sh}=4$ decelerating shocks with different formation depths.  The 4-velocity of the $k_{th}$ shock is
given by $u^k_{sh}(\tau)=10(\tau/\tau^k_0)^{1/2}$, with $\ln\tau^k_0=\ln 2+ k\ln 250/(N_{sh}-1)$ for $k=0,1,2,3$.}
 \end{figure}

A non-uniform dissipation profile will be established if the shock weakens
as it propagates. That happens for example when a thin, fast shell collides
with a slower, much thicker shell. In such a case, the peak is
shifted to much lower energies, and the peak intensity is also reduced (see Fig. \ref{fig:f2}).
Figure \ref{fig:f6} presents the TSED emitted by an ensemble of $N_{sh}=4$ decelerating shocks, having a
velocity profile $u^k_{sh}(\tau)=10(\tau/\tau^k_0)^{1/2}$, with $\ln \tau_0^k$ equally spaced in the range 
$\ln 2\le \ln \tau_0^k\le \ln 200$.  Such a profile can describe
the dynamics of a blast wave propagating in a medium having a density
profile $n_{b}\propto r^{-2}$.  As seen, the TSED in this example has a wiggly shape, with an average slope 
of 1.2 (i.e., $\nu E_{\nu}\propto\nu^{1.2}$).   This wiggly substructure depends more sensitively on the distribution of shock strengths and other details. 
However, it is exaggerated in our calculations for two reasons:
First, our choice of well separated shocks. Second, the assumption that the local spectrum emitted by each shocked shell
is a Wien spectrum.   As explained above, at modest optical depths the local spectrum is expected to be softer, and this should 
lead to a smoothing of the spectrum.  At any rate, when the shock spacing is small enough this substructure completely disappears. 
Wiggles are occasionally seen 
in the sub-thermal component of the prompt spectrum.  Whether theses are instrumental effects or physical effects is unclear 
to us, but if real can be interpreted as resulting from a multiple shock event. 

 \begin{figure}[ht]
\centering
\includegraphics[width=12cm]{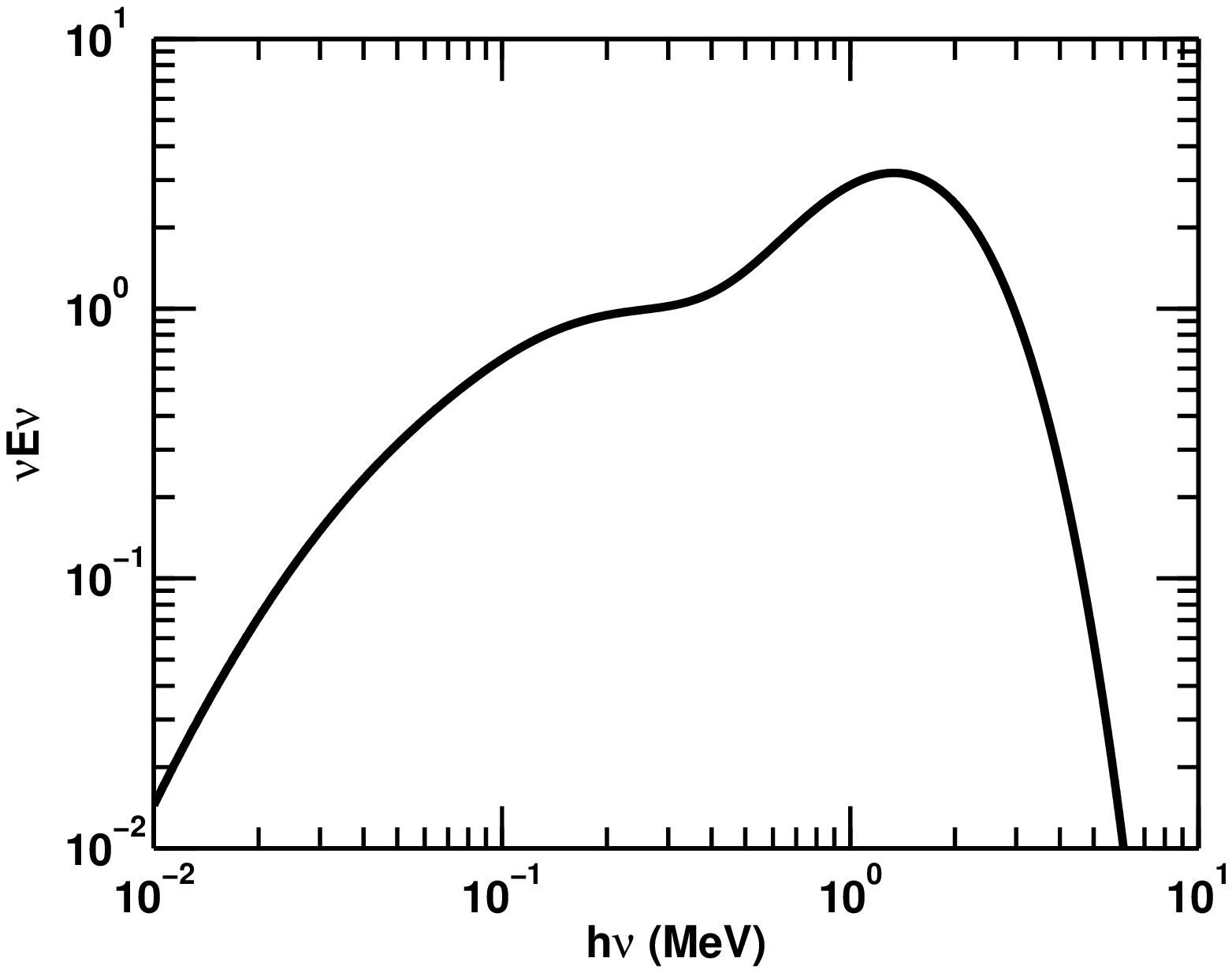}
\caption{\label{fig:f7} An example of a double-peak TSED emitted by a shock train with $u_{sh}^k=0.2, 0.3, 0.4, 0.5$ and $1.3$.}
 \end{figure}

The double-peak SED observed in 
some bursts, notably GRB 110721A and GRB 120323, can also be 
reproduced by the shock train model.   The main (higher energy) peak is attributed to a fast shock, and
the low-energy spectral component, below the second peak, to a sequence of slower shocks.  
An example is displayed in figure \ref{fig:f7}, for a series of shocks with  velocities $u^k_{sh}=0.2,0.3,0.4,0.5$ and $1.3$. 

\section{Summary and conclusions}
We computed the time integrated, spectral energy distribution emitted by a train of radiation mediated shocks that breaks out of a GRB outflow. 
In this paper we considered only the thermal emission by the shocked plasma, that contributes to the portion
of the spectrum below the peak.   As explained elsewhere (LE12), a non-thermal component, extending up to the
KN limit (a few hundred MeV in the observer frame), is expected to be produced via  bulk Comptonization inside the shock transition layer.  Such a component was indeed found in Budnik et al. (2010), who performed detailed calculations of RRMS under the conditions  anticipated
during shock breakout in supernovae, and in the preliminary Monte-Carlo simulations reported in Bromberg et al. (2011).  A full, self-consistent calculation 
of the spectrum produced inside the shock, in the context of the model outlined in this paper, are in progress, and will be reported in a future publication.  Additional contribution
to the non-thermal emission is expected in the post breakout phase, after the shock becomes collisionless.  
The relative importance of this process depends primarily on the fraction of shock energy remaining after breakout.  

Our main conclusion is that the overall shape of the time integrated SED emitted by a shock train with a sufficient 
spread in shock properties, is a broken power law, with a slope $1<\alpha<2$ below the peak (for $\nu E_\nu\propto \nu^\alpha$).   
This conclusion is in line with recent works (Giannois 2012; Vurm et al. 2013;  Beloborodov 2013) showing that, 
quite generally, extended dissipation below the photosphere can naturally produce a Band-like spectrum in the prompt 
phase.  However, in these works the dissipation mechanism is not specified, but rather ad hoc assumptions are made 
about the dissipation profile.  Our calculations are based on a specific dissipation
mechanism, starting from first principles. 

When the number of shocks contributing to the emission is small, the spectrum below the peak may exhibit a wiggly structure.   In particular,
we demonstrated that a double-peak SED, as observed in some bursts, e.g., GRB 110721A and GRB 120323, can be produced
by this model under certain conditions.   Theses effects may depend on details, and need to be explored further using more refined calculations.

This research was supported by a grant from the Israel Science Foundation no. 1277/13

\end{document}